\documentclass[aps,prl,reprint,floatfix,amsmath,amssymb,
superscriptaddress,nofootinbib]{revtex4-1}

\usepackage{graphicx}
\usepackage{epstopdf}
\usepackage{color}
\usepackage{bm}
\usepackage{printlen}
\usepackage{units}

\usepackage[colorlinks=true,citecolor=cyan,linkcolor=magenta]{hyperref}

\usepackage{ulem}

\def\bk{{\bf k}}

\usepackage{nicefrac}

\definecolor{darkgreen}{rgb}{0.0, 0.26, 0.15}

\definecolor{rred}{rgb}{0.77, 0.12, 0.23}

\definecolor{orange}{rgb}{1.0, 0.49, 0.0}

\begin{document}

\title{Landau Levels as a Probe for Band Topology in Graphene Moir\'e Superlattices}
 
\author{QuanSheng Wu}
\email{quansheng.wu@epfl.ch}
\affiliation{Institute of Physics, Ecole Polytechnique F\'{e}d\'{e}rale de Lausanne (EPFL), CH-1015 Lausanne, Switzerland}
\affiliation{National Centre for Computational Design and Discovery of Novel Materials MARVEL, Ecole Polytechnique F\'{e}d\'{e}rale de Lausanne (EPFL), CH-1015 Lausanne, Switzerland}

\author{Jianpeng Liu}
\affiliation{School of Physical Science and Technology, ShanghaiTech University, Shanghai, 200031, China
}
\affiliation{ShanghaiTech laboratory for topological physics, ShanghaiTech University, Shanghai, 200031, China
}

\author{Oleg V. Yazyev} 	
\email{oleg.yazyev@epfl.ch}
\affiliation{Institute of Physics, Ecole Polytechnique F\'{e}d\'{e}rale de Lausanne (EPFL), CH-1015 Lausanne, Switzerland}
\affiliation{National Centre for Computational Design and Discovery of Novel Materials MARVEL, Ecole Polytechnique F\'{e}d\'{e}rale de Lausanne (EPFL), CH-1015 Lausanne, Switzerland}

\date{\today}

\begin{abstract}

We propose Landau levels as a probe for topological character of electronic bands in two-dimensional moir\'e superlattices. We consider two configurations of twisted double bilayer graphene (TDBG) that have very similar band structures, but show different valley Chern numbers of the flat bands. These differences between the AB-AB and AB-BA configurations of TDBG clearly manifest as different Landau level sequences in the Hofstadter butterfly spectra calculated using the tight-binding model. The Landau level sequences are explained from the point of view of the distribution of orbital magnetization in momentum space that is governed by the rotational $C_2$ and time-reversal $\mathcal{T}$ symmetries.
Our results can be readily extended to other twisted graphene multilayers and $h$-BN/graphene heterostructures thus establishing the Hofstadter butterfly spectra as a powerful tool for detecting the non-trivial valley band topology.
\end{abstract}
\maketitle
 
The recent discovery~\cite{Cao2018a,Cao2018b,Lu2019,Chen2019,Burg2019,Shen2020} of correlated insulating phases, unconventional superconductivity, and (quantum~\cite{Serlin2020}) anomalous Hall effect~\cite{Sharpe605,Liu2019,Bultinck2019} in twisted bilayer graphene (TBG) and related moir\'e superlattices
have drawn widespread attention from in theoretical and experimental physics communities. In these twisted graphene multilayers, the width of the four-band manifold around the charge neutrality point (CNP) vanishes at the so-called ``magic'' angle~\cite{Suarez2010,Bistritzer2011}. These flat bands often have non-trivial topology
such as the recently proposed fragile topology~\cite{Song2019,Po2019,Ahn2019}. 
Although the physical mechanisms underlying the observed novel correlated phases are still under debate, the small bandwidth and the non-trivial topology of the relevant bands are certainly pointing to new, interesting physics. However, directly probing the topological properties in experiments is difficult due to their ``hidden'' nature: the topological properties of the two valleys intrinsic to the electronic structure of these systems would cancel each other provided that valley degeneracy is preserved.

In this Letter, we propose Landau levels as such a probe of the topological character of electronic bands in graphene moir\'e superlattices. We illustrate this idea using the example of twisted double bilayer graphene (TDBG), a system constructed by twisting two AB-stacked bilayer graphene (BLG) counterparts placed on top of each other. This more complex four-layer moir\'e heterostructure has recently revealed several novel properties such as the gap opening at large twist angles ~\cite{Haddadi2019,Adak2020,Rickhaus2019,Choi2019,Culchac2020,Chebrolu2019} and two types of stacking configurations that have distinct topological properties~\cite{Liu2019,Chebrolu2019}. Moreover, the band structure and topological properties of TDBG can be controlled by applying external electrical fields ~\cite{Liu2019,Chebrolu2019,Shen2020,Lee2019,Koshino2019}, and could lead to quantum anomalous Hall effect when correlation effects are taken into account~\cite{Liu2019b}. 

\begin{figure}[b]
    \centering
    \includegraphics[width=.5\textwidth]{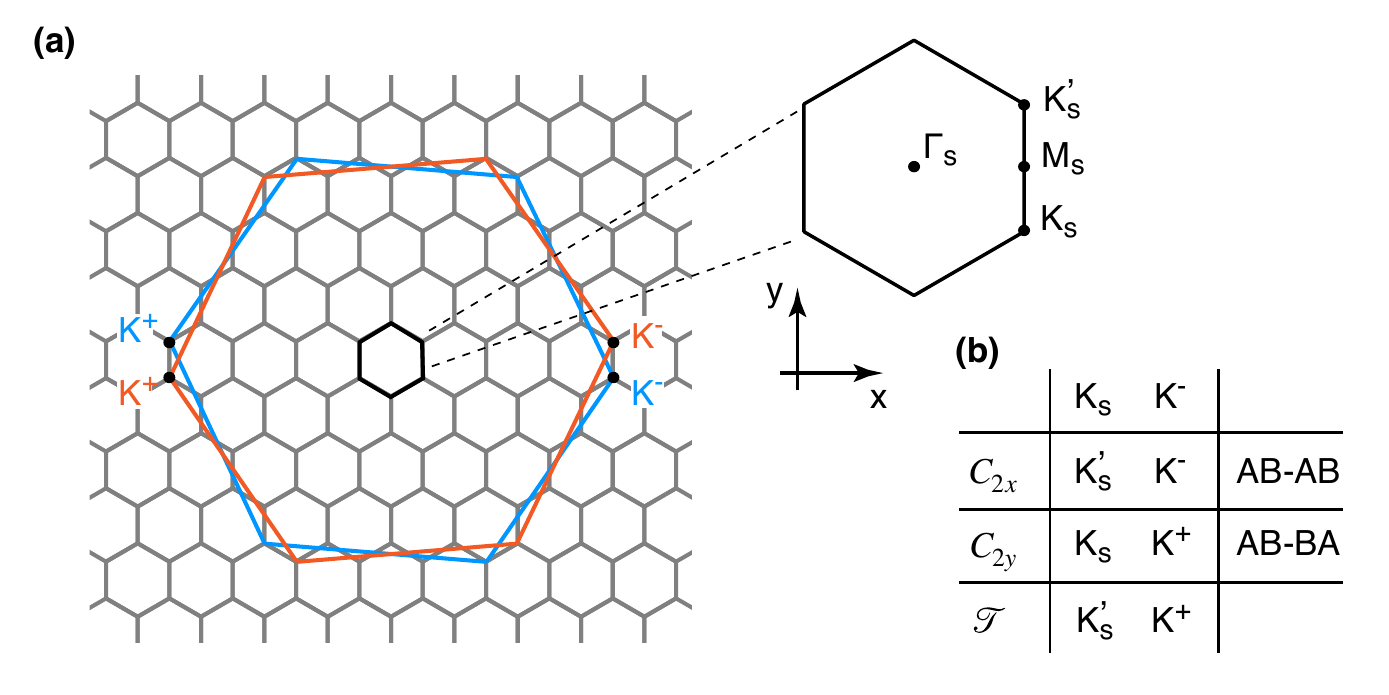}
    \caption{(a) Brillouin zones of the two BLG components (orange and blue for top and botton bilayers, respectively) and moir\'e supercell (grey hexagons). (b) Change of valley momenta under rotational ($C_{2x}$, $C_{2y}$) and time-reversal ($\mathcal{T}$) symmetry operations. }
    \label{fig:tdbg_bz}
\end{figure}

Two distinct configurations of TDBG referred to as AB-AB and AB-BA are related to each other by rotating the BLG counterparts by 180$^\circ$ with respect to each other. Both belong to the $D_3$ symmetry group, but differ by having the $C_{2x}$ and $C_{2y}$ symmetries, respectively. 
The band structures of the AB-AB and AB-BA configurations were found to be similar~\cite{Culchac2020,Koshino2019}, but 
the above-mentioned symmetry differences result in different band topologies.
The $C_{2x}$ symmetry requires the Chern number for each valley to be vanishing, while $C_{2y}$ doesn't. The time-reversal symmetry requires the Chern numbers of the two valleys are opposite.
Hence, the AB-AB configuration of TDBG has trivial valley Chern numbers, while the AB-BA configuration is topologically nontrivial. The Chern number is the integral of Berry curvature that affects the Landau level (LL) spectrum when magnetic field is applied~\cite{Chang_2008,PhysRevB.59.14915}. We show that the LL spectra of the AB-AB and AB-BA configurations of TDBG are  dramatically different, which allows to discriminate them despite their virtually indistinguishable band structures.

The Hofstadter butterfly (HB) theoretically proposed in 1976 is a self-similar recursive Landau level spectrum of a system subject to both magnetic field and  periodic potential~\cite{Hofstadter1976}. Its experimental observation requires that the characteristic length of magnetic field is comparable to the lattice constant (magnetic field of 1~Tesla corresponds to the characteristic length of 25.7~nm). Lattice constants that are sufficiently large for observing the HB spectra can be achieved in moir\'e superlattices realized by stacking two periodic lattices with different lattice constants, as first realized in the the graphene/h-BN system~\cite{Dean2013}, or by twisting them with respect to each other.
In the latter case, the lattice constant of the moir\'e superlattice can be controlled by the twist angle, making it a versatile platform for studying the HB physics. 

\begin{figure}[t]
    \centering
    \includegraphics[width=.5\textwidth]{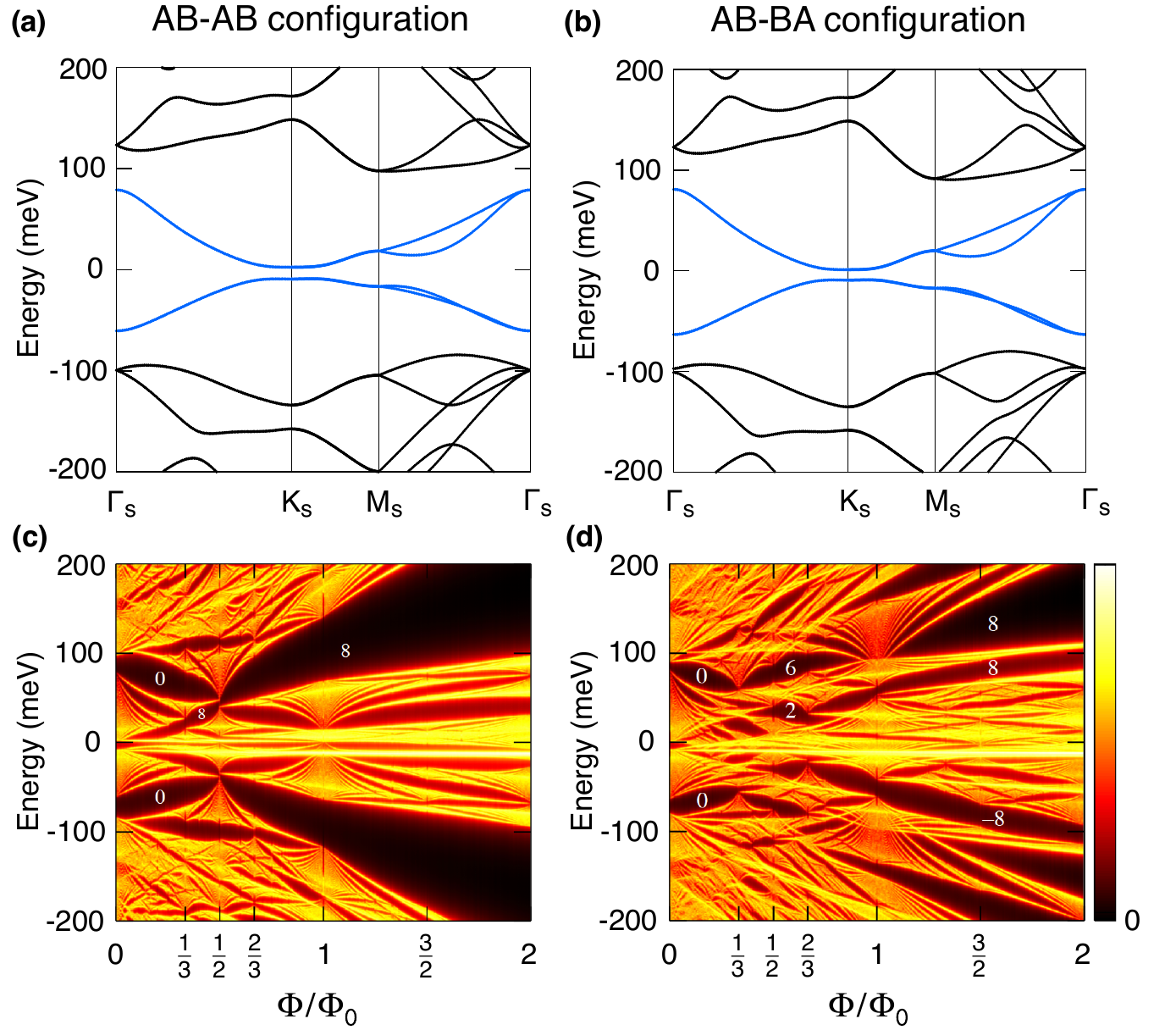}
    \caption{(a,b) Band structures and (c,d) Hofstadter butterfly spectra of the AB-AB and AB-BA configurations of TDBG, respectively, characterized by twist angle $\theta = 1.89^\circ$. The flat band manifold is shown in blue. 
    The numbers in the HB spectra indicate the 
    Chern numbers of the LLs gaps. }
    \label{fig:HB_m17}
\end{figure}

The HB spectrum and LLs of TBG close to the magic angle have recently been investigated in several works~\cite{Lian2018,Hejazi2019,Zhang2019}. Lian {\it et al.} studied the HB of TBG, and found that the HB of the flat-band manifold is generically connected with the remote bands since the flat bands have non-trivial fragile topology~\cite{Lian2018}. Zhang {\it et al.} found that the degeneracy of the LLs would be lifted when the crystal symmetry is broken~\cite{Zhang2019}. 
In our work, we show that the distribution of orbital magnetization in momentum space can lift the LL degeneracy, and that the LL splittings are crucially dependent on the stacking configuration and band topology of the TDBG system.

The tight-binding (TB) Hamiltonian in presence of a magnetic field is obtained by adding phase factors $\phi_{ij}$ to the corresponding hopping integrals, a procedure known as the Peierls substitution,
\begin{eqnarray}
    \hat{H}=\sum_i \epsilon_i c^{\dag}_i c_i +\sum_{<i,j>}t_{ij}e^{i\phi_{ij}}c^{\dag}_i c_j , \label{eqn:H}\\
    \phi_{ij}=\frac{2\pi}{\Phi_0}\int_{\mathbf{r_i}}^{\mathbf{r_j}}{\mathbf{A(r)}}\cdot d\mathbf{r} ,
\end{eqnarray}
where $\epsilon_i$ is the onsite energy, $\mathbf{r_i}$ is the atom's position, $\mathbf{A(r)}$ is a vector potential and $\Phi_0=h/e$ is the magnetic flux quantum with $e$ being the electron charge and $h$ the Planck constant. 
The TB parameters $\epsilon_i$ and $t_{ij}$ are deduced from first-principles calculations and take into account the lattice relaxation effects obtained using atomistic classical force field simulations. Applied electric field and intrinsic polarization effects were not considered in the reported calculations. Further details of the methodology can be found in Refs.~\onlinecite{Gargiulo2018,Haddadi2019}. 
The phase factor $\phi_{ij}$ is not periodic  modulo $2\pi$ in the usual Landau gauge $\mathbf{A}=B x \hat{e}_y$ when $\mathbf{r}_i$ and $\mathbf{r}_j$ are not nearest neighbours. In order to cope with this problem, we adopt the periodic Landau gauge introduced by Nemec and Cuniberti~\cite{Nemec2007} and further used by Hasegawa and Kohmoto~\cite{Hasegawa2013} to study TBG. This periodic Landau gauge is defined as
\begin{eqnarray}
    \mathbf{A(r)}=\frac{\Phi}{2\pi}\left((\xi_1-\lfloor\xi_1\rfloor)\mathbf{K_2}-\xi_2\sum_{n=-\infty}^{\infty}\delta(\xi_1-n+\epsilon)\mathbf{K_1}\right),
    \nonumber\\
\end{eqnarray}
where ($\xi_1,\xi_2$) are the oblique coordinates defined by $\mathbf{r}=\xi_1 \mathbf{R_1}+\xi_2 \mathbf{R_2}$ with $\mathbf{R_1},\mathbf{R_2}$ being the primitive vectors of the moir\'e unit cell, $\mathbf{K_1}$, $\mathbf{K_2}$ are the corresponding reciprocal lattice vectors, $\epsilon$ is a positive infinitesimal and $\lfloor x\rfloor$ is the floor function defined as largest integer not greater than $x$. $\Phi$ is the magnetic flux through the moir\'e unit cell defined as 
\begin{eqnarray}
\Phi=BS=\frac{p}{q}\Phi_0 ,
\end{eqnarray}
where $S$ is the area of the moir\'e unit cell, $p$ and $q$ are co-prime integers. The size of magnetic supercell is $q$ times the moir\'e unit cell along the $\mathbf{R}_2$ direction. The HB and LLs spectra, represented by the local density of states, are obtained by numerically solving Eqn.~(\ref{eqn:H}) using the Lanczos recursive method as implemented in the WannierTools open-source software package~\cite{Wu2017}. 

Without loss of generality, we will focus on TDBG with twist angle $\theta=1.89^\circ$, for which we set $q = 500$ in our calculations. As shown in Figs.~\ref{fig:HB_m17}a,b, the band structures of the AB-AB and AB-BA configurations are practically indistinguishable as far as the flat-band manifold is concerned. Figs.~\ref{fig:HB_m17}c,d show the HB spectra of these two TDBG configurations. It is evident that despite very similar band structures, the AB-AB and AB-BA configurations have very different HB spectra as well as Chern numbers associated with the LL gaps. The LLs of the flat bands are connected with the LLs originating from higher energy bands in both cases, which is observed also for smaller twist angles. Lian {\it et al.}~\cite{Lian2018} attributed this to the nontrivial fragile topology of TBG. However, we note that no fragile topology and no valley Chern numbers characterize the AB-AB configuration of TDBG.  

\begin{figure*}[!htbp]
    \centering
    \includegraphics[width=\textwidth]{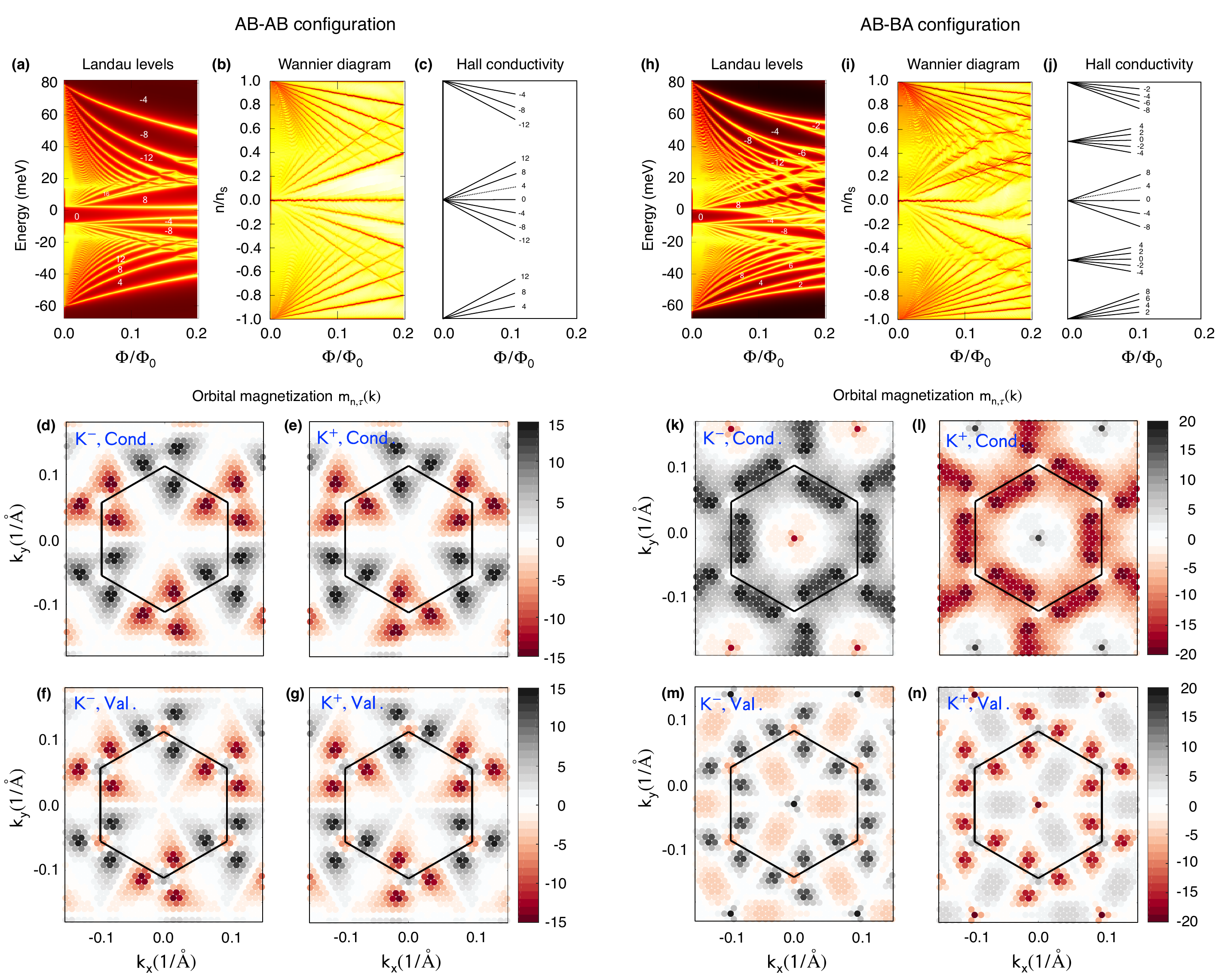}
    \caption{Landau levels, Wannier diagrams, Hall conductivity and orbital magnetization plots for the flat-band manifold in the AB-AB (left) and AB-BA (right) configurations of TDBG at twist angle $\theta=1.89^\circ$. (a,h) The LL spectra as a function of magnetic flux per moir\'e unit cell. The valley Chern numbers of the LL gaps are indicated. (b,i) Normalized charge-carrier density per moir\'e unit cell as a function of magnetic field flux. The linear trends correspond to the gaps, hence the LL filling factors can be deduced from the slopes of these lines. (c,j) Quantized Hall conductivity of the Landau fans. Panels (d-g) and (k-n) show the orbital magnetization $m_{n,\tau}(\mathbf{k})$ in units of $\mu_B$, where $n$ is the band index representing conduction or valence bands and $\tau$ is the graphene valley index $K^+$ or $K^-$. }
    \label{fig:HB_WD_TDBG-m17}
\end{figure*}

A convenient way for observing the HB in experiments relates to the Wannier diagrams (WDs) obtained by plotting the Hofstadter energy spectrum as integrated charge-carrier density $n$ versus magnetic field $B$ or magnetic flux $\Phi$~\cite{Wannier1978}. WDs show that all spectral gaps are constrained to linear trends in the density-field diagrams. This can be described by a simple Diophantine relation
\begin{eqnarray}
n/n_s=t \Phi/\Phi_0 +s ,
\end{eqnarray}
where $n/n_s$ and $\Phi/\Phi_0$ are the normalized carrier density and magnetic flux, respectively, and $s$ and $t$ are integer numbers. Here, $n/n_s$ represents the Bloch band filling fraction. The first quantum number $t$ is related to the Hall conductivity $\sigma_{xy}$ associated with each minigap in the fractal spectrum. $\sigma_{xy}$ is quantized according to the relation $\sigma_{xy}= 4te^2/h$, where factor 4 originates from the valley and spin degeneracies. The second quantum number $s$ corresponds to the Bloch band filling index in the fractal spectrum.

In the limit of weak out-of-plane uniform fields $\mathbf{B}=(0,0,B)$, the evolution of energy bands can be treated perturbatively as \cite{Chang1996,PhysRevB.59.14915,Chang_2008,Sun2020} 
\begin{equation}
    \varepsilon_{n,\sigma,\tau}({\mathbf{k},B})=\varepsilon_{n,\tau}({\mathbf k})+\mu_Bg{\mathbf{\sigma}} {B}+m_{n,\tau}({\mathbf k}) { B},\label{eqn:Ek_B}
\end{equation}
where $\mathbf{\sigma}$ is the electron spin operator assuming $\pm$1/2 values for up and down spins, respectively, and $\mathbf{\tau}=\pm1$ is the valley index. The valley orbital magnetization is defined as 
\begin{equation}
\small{}
    m_{n,\tau}(\mathbf k)=-\mu_B\frac{2m_e}{\hbar^2}\mathrm{Im}\sum_{l\neq n}\frac{\langle n,\tau|\partial_{k_x}\mathcal{H}_{\tau}|l,\tau\rangle \langle l,\tau|\partial_{k_y}\mathcal{H}_{\tau}|n,\tau\rangle}{\varepsilon_{n,\tau,\mathbf{k}}-\varepsilon_{l,\tau,\mathbf{k}}} .
\end{equation}
There are two contributions to the energy due to magnetic field. The first contribution originating from the Zeeman effect of electron spin is neglected throughout this paper for simplicity. The second contribution is related to the orbital magnetization contribution $m_{n,\tau}(\mathbf k)$. 

The LL spectra, Wannier diagrams and the distribution of orbital magnetization in momentum space for the the flat-band manifold of the AB-AB and AB-BA configurations of TDBG at $\theta=1.89^\circ$ in a low-field range are presented in
 Fig.~\ref{fig:HB_WD_TDBG-m17}. 
In the case of Bernal (AB-stacked) BLG, the sequence of the Hall conductivity values $\sigma_{xy} =\pm4, \pm8, \pm12, ...$~$e^2/h$~\cite{Novoselov2006} with the increment of 4~$e^2/h$ is related to the combination of the spin and (bilayer graphene) valley degeneracies.
In TDBG, the moir\'e valley degeneracy adds to the above degeneraciers increasing the increment of the Hall conductivity sequence to 8~$e^2/h$. In our calculations, however, we observe the 4~$e^2/h$ increment close to the CNP for both the AB-AB and AB-BA configurations of TDBG (Figs.~\ref{fig:HB_WD_TDBG-m17}c,j). This implies that one of three degeneracy flavors is lifted under applied magnetic field. Due to the neglected
Zeeman effect term, either bilayer graphene valley or moir\'e valley degeneracies are expected to be lifted by magnetic field. In order to clarify this issue, we consider the transformations of orbital magnetization $m_{n,\tau}(\mathbf k)$ under the $C_{2x}$, $C_{2y}$ and $T$ symmetries:
\begin{eqnarray}
&T :& m_{n}(\bk) =-m_n(-\bk),\\
&C_{2x} :& m_{n}(k_x,k_y) =-m_n(k_x,-k_y),\\
&C_{2y} :&  m_{n}(k_x,k_y) =-m_n(-k_x,k_y).
\end{eqnarray}

In the AB-AB configuration of TDBG, the $C_{2x}$ symmetry operation exchanges moir\'e valleys $K_s$ and $K_s'$ while keeping the bilayer graphene valleys $K^+$ and $K^-$ unchanged (Fig.~\ref{fig:tdbg_bz}).
Eventually, the orbital magnetization $m_{n,\tau}(\mathbf k)$ is the same for the two bilayer graphene valleys while it is opposite in the two moir\'e valleys. The orbital magnetization $m_{n,\tau}(\mathbf k)$ of the conduction and valence bands for the two valleys, calculated using the continuum model Hamiltonian described in Ref.~\onlinecite{Liu2019}, is shown in Fig.~\ref{fig:HB_WD_TDBG-m17}d-g. The results are fully consistent with our symmetry analysis. The Landau levels at the CNP originate from the energy bands at the two moir\'e valleys $K_s$ and $K_s'$.
According to Eqn.~(\ref{eqn:Ek_B}), the LLs originating from moir\'e valleys $K_s$ and $K_s'$ are no longer degenerate due to their opposite orbital magnetization $m_{n,\tau}(\mathbf k)$, while the LLs of the two bilayer graphene valleys preserve the degeneracy due to the same orbital magnetization. To support this argument, let us consider the lowest LL of the valence and conduction bands shown in Figs.~\ref{fig:HB_WD_TDBG-m17}a,b. The large splitting of the lowest LLs originating from the valence band contrasts with essentially no splitting for the conduction band LLs. This can be explained by the fact that
$m_{n,\tau}(\mathbf k)$ of the valence band at $K_s$ and $K_s'$ is about $\pm6.5\mu_B$ 
while that of the conduction band is zero. 
To provide a rough estimate, the energy splitting at $\Phi/\Phi_0=0.1$ (corresponds to $B\approx9$~T) 
assuming a orbital magnetization of 6.5$\mu_B$ is {\it ca.} 3.2~meV which is comparable to the lowest LL splitting of the valence band shown in Fig.~\ref{fig:HB_WD_TDBG-m17}a. Note that the orbital magnetization of the conduction band at $K_s$ and $K_s'$ is close to zero as shown in Figs.~\ref{fig:HB_WD_TDBG-m17}d,e. Eventually, the LL splitting of the conduction band close to CNP is much weaker than that of the valence band. For this reason, the splitting of the lowest LL of the conduction band at CNP is missing, which manifests in apparent absence of $\sigma_{xy} = 4$~$e^2/h$ from the Hall conductivity sequence (Fig.~\ref{fig:HB_WD_TDBG-m17}c). The same scenario is also observed for the AB-BA configuration of TDBG discussed below. The LLs at $n/n_s=\pm 1$ originate from the $\Gamma_s$ point where the orbital magnetization of the conduction and valence bands is zero due to the symmetry constrain. 
Eventually, as shown in Figs.~\ref{fig:HB_WD_TDBG-m17}a-c, the sequence of the LLs at $n/n_s=\pm 1$ is 0, $\pm 4$, $\pm 8$,... with increment of 4 originating from the combination of spin and bilayer graphene valley degeneracies. 

In the AB-BA configuration of TDBG, the $C_{2y}$ symmetry exchanges bilayer graphene valleys $K^-$ and $K^+$ while keeping the moir\'e valleys unchanged. 
In this case, the orbital magnetization $m_{n,\tau}(\mathbf k)$ shown in Figs.~\ref{fig:HB_WD_TDBG-m17}k-n is the same for the two moir\'e valleys, while it is opposite for the two bilayer  graphene valleys. The latter indicates that the bilayer  graphene valley degeneracy of LLs is lifted under magnetic field, as supported by Figs.~\ref{fig:HB_WD_TDBG-m17}h-j. The Hall conductivity sequence at CNP $n/n_s=0$ is $\sigma_{xy}=0,\pm4, \pm8, ...$~$e^2/h$, {\it i.e.}
the same as for the AB-AB configuration. However, at $n/n_s=\pm1$ the Hall conductivity sequence $\sigma_{xy}=0,\pm2, \pm4,...$~$e^2/h$ with increment of 2~$e^2/h$ is different from that of the AB-AB configuration. Furthermore, another Landau fan at half-filling $n/n_s=1/2$ can be observed, while it is absent in the case of AB-AB configuration of TDBG. This Landau fan at $n/n_s=1/2$ appears when the degeneracy is lifted in the whole BZ. 

In conclusion, through large-scale numerical calculations based on the atomistic tight-binding model and symmetry analysis, we have investigated the LL spectra of two configurations of TDBG with the same value of twist angle. It was found that the LL sequences close to the CNP of both systems are very similar although their origin is different, while the LL sequences at $n/n_s=\pm1$ and $n/n_s=\pm 1/2$ of both systems are very different. These similarities and differences are caused by the momentum-space distribution of orbital magnetization $m_{n,\tau}({\bk})$  subject to symmetries. These considerations can be readily generalized to a broader class of moir\'e superlattice systems, such as other twisted graphene multilayers and $h$-BN/graphene heterostructures, characterized by flat bands with non-trivial valley Chern numbers.
Our results thus suggest Landau levels as a versatile experimental probe for the ``hidden'' topological character of bands in two-dimensional moir\'e systems.  

Q.W. and O.V.Y. acknowledge support by NCCR Marvel. Computations were performed at the Swiss National Supercomputing Centre (CSCS) under projects Nos. s832 and s1008 and the facilities of Scientific IT and Application Support Center of EPFL.

\bibliography{refs}{}
\end{document}